\documentclass[preprint]{elsarticle}
\usepackage{graphicx,url,hyperref,microtype,subcaption,amsmath,graphbox}
\usepackage{latexsym,courier,upgreek,xcolor,rotating,soul}
\usepackage{algorithmic}
\usepackage[displaymath,mathlines]{lineno}
\usepackage[margin=2cm]{geometry}
\hypersetup{colorlinks,citecolor=blue,linkcolor=red,urlcolor=green}
\modulolinenumbers[5]
\biboptions{numbers,sort&compress}

\begin{document}

\begin{frontmatter}
  
  \title{Crack Localization and the Interplay between Stress
    Enhancement and Thermal Noise}
  
  \author[add1,add2]{Santanu Sinha}
  \ead{santanu@csrc.ac.cn}
  
  \author[add2]{Subhadeep Roy}
  \ead{subhadeep.roy@ntnu.no}
  
  \author[add2]{Alex Hansen}
  \ead{alex.hansen@ntnu.no}
  
  \address[add1]{Beijing Computational Science Research Center, 10
    East Xibeiwang Road, Haidian District, Beijing 100193, China.}
  
  \address[add2]{PoreLab, Department of Physics, Norwegian University
    of Science and Technology, N-7491 Trondheim, Norway.}
  
  \begin{abstract}
    We study the competition between thermal fluctuations and stress
    enhancement in the failure process of a disordered system by using
    a local load sharing fiber bundle model. The thermal noise is
    introduced by defining a failure probability that constitutes the
    temperature and elastic energy of the fibers. We observe that at a
    finite temperature and low disorder strength, the failure process,
    which nucleate in the absence of any thermal fluctuation, becomes
    spatially uncorrelated when the applied stress is sufficiently
    low. The dynamics of the model in this limit lies closely to the
    universality class of ordinary percolation. When applied stress is
    increased beyond a threshold value, localized fractures appear in
    the system that grow with time. We identify the boundary between
    the localized and random failure process in the space of
    temperature and applied stress, and find that the threshold of
    stress corresponding to the onset of localized crack growth
    increases with the increase of temperature.
  \end{abstract}

  \date{\today}
  
  
\end{frontmatter}


\makeatletter
\def\ps@pprintTitle{%
 \let\@oddhead\@empty
 \let\@evenhead\@empty
 \def\@oddfoot{}%
 \let\@evenfoot\@oddfoot}
\makeatother

\section{Introduction}
Growth of fractures under external stress in heterogeneous materials,
such as concrete or fiber-reinforced composites, depends on the
interplay between material disorder and local stress concentration
throughout the failure process. The heterogenities compete with the
local stress enhancement preventing the growth of a single unstable
crack until the stress exceeds a critical value, beyond which it
undergoes a catastrophic failure \cite{hr14}.  In addition to material
heterogeneities, there is also the disorder induced by thermal
fluctuations.  These combined with an applied stress or strain can
cause failure with time even if the applied stress is below the
critical value \cite{l93}. This phenomenon is known as creep
failure. The study of creep failure is an essential but notoriously
difficult subject that has important engineering applications.  Of
special importance is the statistics connected to the time elapsed
until creep failure occurs, the creep lifetime \cite{wb12,chhp20}.

The underlying physical mechanisms behind creep are the accumulation
of plastic strain and damage in the system. In order to model the
creep failure in disordered systems, mechanisms for such aging
processes are needed to be taken into account in addition to the
structural heterogeneities. However, the fracture of heterogeneous
solids is yet hard to handle with the elasticity theory and therefore
there have been some alternative approaches. The Voight model
\cite{v89} for precursory strain has been explored by Main \cite{m00}
in order to understand the sub-critical crack growth dynamics. The
model can reproduce the time-dependent strain rate observed in
experiments \cite{nhg05}. On the other hand, time-dependent failure
has been investigated widely \cite{cgs01, scg01, pcs02, ss05, hmk02,
  hkh01, hkh02, kmh03, pc03, dk13, pcc13} using a simple and intuitive
model for disordered solids, known as the fiber bundle model
\cite{p26, d45, hr90, cb97, phc10, hhp15, brc15}. Here, the thermal
fluctuations are incorporated in terms of noise or probabilistic time
evolution in order to include the effect of temperature during
failure. These studies in fiber bundle model appear to be successful
in explaining the time-dependent strain rate as well as the
temperature and the stress (or strain) dependencies in the creep
lifetime \cite{cgs01, scg01, pcs02, ss05}. Hidalgo et al., however,
obtained somewhat different results \cite{hkh02, kmh03} for the
temperature and stress dependence of creep lifetime from those in the
above-mentioned stochastic models \cite{cgs01, scg01, pcs02, ss05} by
introducing a time evolution equation for the strain of each
constituent (i.e., fiber) based on the Kelvin-Voigt rheology. Danku
and Kun consider a damage accumulation process in each fiber by
introducing a damage variable and the time evolution equation
\cite{dk13}. Interestingly, fiber bundle model at its simplest form
can exhibit creep-like behaviors even in the absence of any thermal
fluctuations, damage variables, or rheological constitutive laws
\cite{rbr19, rh18}. In particular, Roy and Hatano assumed the time
evolution in a simple fiber bundle model and derived the
time-dependent strain rate \cite{rh18}.

Here we study the role of thermal fluctuations in the failure process
of a disordered system by using a local-load sharing (LLS) fiber
bundle model where the material disorder and local stress enhancements
compete. We introduce thermal noise in the model by a probabilistic
algorithm that is based on the elastic energy and the breaking energy
of a fiber. This creates time-dependent fluctuations in the model in
addition to the quenched system disorder.  The interplay between these
two types of disorders and the local stress enhancements creates
non-trivial dynamics in the failure process. More specifically, we
show that the introduction of thermal noise makes the failure process
non-localized even at low system disorder under low stress, which
otherwise would be localized in the absence of thermal
fluctuations. We characterize the non-localized growth regime by
measuring the geometrical properties of the cracks and we show that
they belong to the percolation universality class. We also identify
the boundary between the localized and non-localized regimes in the
temperature-applied stress plane. We present the model in Section
\ref{secModel} and the numerical results in Section \ref{secRes} where
we also discuss the results. We conclude in Section \ref{secCon}.

\section{Model description}
\label{secModel}
The fiber bundle model consists of $N$ Hookean springs or {\em fibers}
placed between two clamps under an external force $F$ carried by the
fibers. The extension $x_i$ of a fiber $i$ under a force $f_i$ follows
the relation $f_i=\kappa x_i$. All the fibers have the same elastic
constant $\kappa$.  Each fiber is assigned a maximum extension
$\epsilon_i$.  If this value is exceeded, that particular fiber
fails. The distribution of the thresholds $\epsilon_i$ among the
fibers models the heterogeneity of the material. When a fiber fails,
the load it was carrying is distributed among the surviving fibers
according to a load-sharing scheme that models the way the forces are
distributed among the surviving fibers. If the load carried by the
failed fibers is distributed uniformly over all the surviving fibers,
we have the {\it Equal Load Sharing\/} (ELS) scheme. With this scheme,
there is no local stress enhancement in the model. On the other hand,
if the load carried by the failing fiber is distributed evenly among
the fibers bordering the cluster of broken fibers to which the failed
fiber belong, we are dealing with the {\em Local Load Sharing} (LLS)
scheme. Here local stress enhancement competes with the local
heterogeneity. The ELS fiber bundle model was initially introduced by
Peirce \cite{p26} to model the strength of yarn and since Daniels'
paper \cite{d45}, the model caught on in the mechanics
community. Sornette introduced the ELS fiber bundle model to the
statistical physics community in 1992 \cite{s92} which lead the
community to explore the rich avalanche statistics
\cite{hh92,hh94,zd94,khh97} and analytical tractability of the ELS
model. The local load-sharing (LLS) fiber bundle model was introduced
by Harlow and Phoenix \cite{hp78,hp91} as a one-dimensional array of
fibers. There are also intermediate models between ELS and LLS models,
e.g.\ the model by Hidalgo et al. \cite{hmk02}, where the load of the
failing fiber is distributed according to a power law in the distance
from the failed fiber.  Another is the soft clamp model
\cite{bhs02,sgh12,gsh13,gsh14}, where the infinitely stiff clamps are
replaced by clamps with finite elastic constant causing the load of
the failing fiber for be distributed among the surviving in accordance
with the elastic response of the soft clamps.

Here we consider an LLS fiber bundle model that is based on a history
independent redistribution scheme \cite{skh15} so that the complete
stress field at any instance can be calculated from the present
arrangement of intact and broken fibers without knowing any
information about order in which the fibers failed. For this, we
define a {\it crack,\/} which is a cluster of $s$ failed fibers
defined as in percolation theory \cite{s79,sa92}. The {\it
  perimeter\/} of the crack is the set of $h$ intact fibers that are
nearest neighbors to the failed fibers in that crack. These nearest
neighbors define the {\it hull\/} of the cluster \cite{ga87}. The
force on an intact fiber $i$ at any instance is then calculated by
\begin{equation}
  \displaystyle
	f_i = f\left(1 + \sum_{J(i)}\frac{s_{J(i)}}{h_{J(i)}}\right)\;,
  \label{eqLLS}
\end{equation}
where $f=F/N$, the force per fiber. The summation runs over all cracks
$J(i)$ that are neighbors to fiber $i$.

Here we study the fiber bundle at a temperature $T$ subjected to a
constant external stress $F$ that is less than the critical breaking
stress for the bundle. In order to do this, we introduce changes to
the LLS model.  Due to thermal noise, a fiber $i$ may fail even if the
force on it fulfills $f_i<\kappa \epsilon_i$. The elastic energy of a
fiber at extension $x_i$ is $\kappa x_i^2/2$ and the elastic energy at
failure is $\kappa \epsilon_i^2/2$. This energy is dissipated, i.e.,
lost as elastic energy at failure. We introduce a discrete time
variable $t$ to be defined below. We define a failure probability
\begin{equation}
  \label{eqProb}
  \displaystyle
  P_i(t,T) = \exp\left[{-\frac{\kappa(\epsilon_i^2-f_i^2)}{2k_BT}}\right]\;,
\end{equation}
where $k_B$ is the Boltzmann constant. In the following we set
$\kappa/k_B=1$ for simplicity. The simulation starts at $t=0$ with all
fibers intact and each carrying a finite load $f_i=f$. The failure
probability $P_i$ is then calculated for each fiber $i$. We then
generate a random number $r_i$ uniformly distributed between $0$ and
$1$ for each fiber $i$, which we compare to $P_i$. All the fibers for
which $r_i<P_i$ fails. The forces are then redistributed according to
Eq. \ref{eqLLS} and time $t$ is increased by 1. This procedure is
repeated until all fibers have failed. Note that the failure
probability $P_i$ is an increasing function of the temperature $T$ as
well as of the local stress $f_i$ which also increases with time due
to the failures. When $f_i > \epsilon_i$, $P_i$ is always greater than
one for any temperature $T$ and the fiber always breaks. Furthermore,
as the temperature is changed towards zero ($T\to 0$), the failure
probability $P_i$ is approached towards a step function from $0$ to
$1$ at $f_i=\epsilon_i$. This means, at $T=0$, there is no
contribution from the thermal noise to the breaking probability and a
fiber only breaks due to the local stress. This leads the model to
approach the conventional LLS model at $T=0$ \cite{srh20}. With
increasing temperature, values of $P_i$ increases throughout the range
of $f_i$ and becomes less dependent on how $f_i$ is close to the
threshold $\epsilon_i$. Recently, a similar probabilistic failure
process was adopted to explore the creep failure in ELS fiber bundle
model \cite{rh20}.

For the disorder in the failure thresholds $\epsilon_i$, we consider a
distribution that has only one control parameter. The thresholds in
this distribution are generated by calling a random number ($r_i$)
over the unit interval and raising it to a power $D$, therefore
$\epsilon_i=r_i^D$. This corresponds to the cumulative distribution
\cite{hhr91,sh19},
\begin{equation}
  \displaystyle
  P(\epsilon) =
  \begin{cases}
    \epsilon^{1/|D|}\;,    \epsilon\in[0,1]       & \text{when} \;  D>0\;, \\
    1-\epsilon^{-1/|D|}\;, \epsilon\in[1,\infty)  & \text{when} \;  D<0\;.
  \end{cases}
  \label{eqPx}
\end{equation}
The strength of disorder in this distribution is controlled by the
value of $|D|$. Moreover, $D>0$ and $D<0$ respectively correspond to
the distributions with power law tails towards weaker and stronger
fibers. The interplay between such disorder and local stress
enhancement in LLS fiber bundle model in the absence of any thermal
noise was explored recently \cite{srh20}. There, phase transitions
from a localized fracture regime to random fractures were observed
while increasing the disorder. Interestingly, the transition for $D<0$
was of first order and for $D>0$ was of second order. In the present
work, we investigate the effect of thermal noise on the failure
dynamics and study its effect on crack localization, and we limit our
simulations in the low disorder limit at $D=0.02$, where the failure
was localized in the absence of temperature. However, as we will see
in the next section that the presence of temperature makes the
fracture growth of this regime non-localized when the stress is
low. We present our numerical results in the following.

\begin{figure}[t]
  \centerline{\hfill\includegraphics[width=0.4\textwidth,clip]{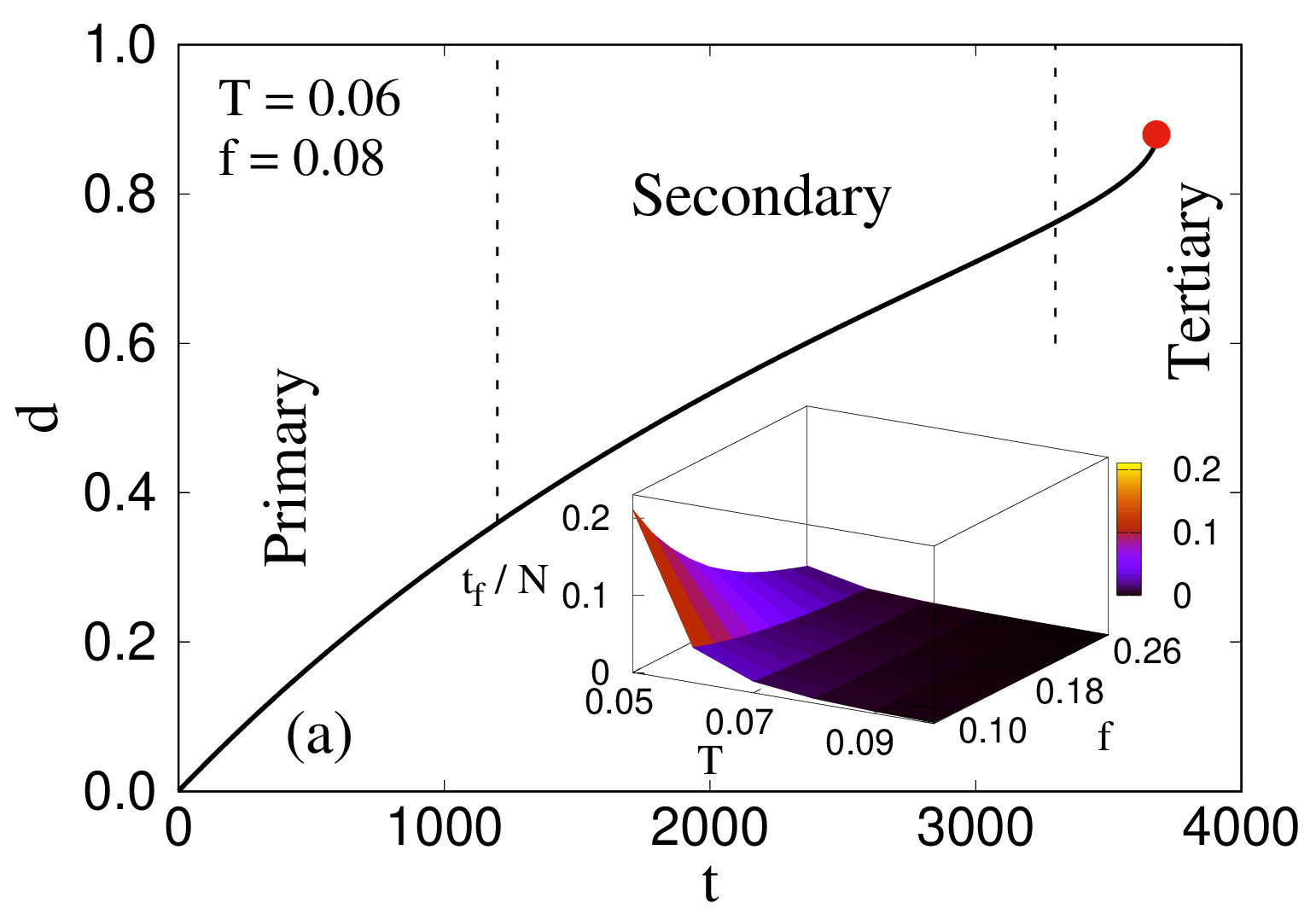}\hfill}
  \caption{\label{figCreep} The variation of damage $d$ with time $t$
    for $T=0.06$ and $f=0.08$ showing three regimes of creep like
    failure, a primary regime a secondary regime and a tertiary
    regime. The red dot shows the final failure point. The
    $x$-coordinate of the point corresponds to the failure time
    $t_{\rm f}$. The inset of the figure shows how $t_{\rm f}$
    decreases with the increase of temperature and applied stress.}
\end{figure}

\section{Results}
\label{secRes}
We consider a bundle of $N=256\times 256$ fibers in two dimensions
(2D) with periodic boundary conditions. The results are averaged over
$1000$ realizations of samples. Simulations are performed for
different values of temperature ($T$) and external stress ($f=F/N$)
which are kept constant throughout a simulation. The thermal noise
initiates a creep failure process to break the system over time even
if the applied stress is less than the critical value. Depending on
the thermal noise and local stress enhancements, many or few fibers
may fail simultaneously. The growth of damage in this process is
illustrated in Fig. \ref{figCreep} where we plot the damage $d$ ---
defined as the number of failed fibers divided by $N$ --- as a
function of $t$ for a stress $f=0.08$, which is lower than the
critical stress. The plot shows three regimes of failure process, a
primary regime where the rate of damage decreases with time, a
secondary regime where the damage rate is almost constant, and a
tertiary regime where the damage increases rapidly until the whole
bundle fails. This is a qualitative characteristics that is generally
observed during a creep failure \cite{hhp15}. The red dot at the end
of the curve indicates the failure time $t_{\rm f}$ which depends on
the temperature as well as on the applied stress. This is shown in the
inset, where $t_{\rm f}$ decreases with increase of both $T$ and $f$,
as a higher stress or temperature generates higher probability of
failure at each time step, making the model go faster towards the
global failure.

\subsection{\bf Localized vs non-localized fracture growth}
\begin{figure}[t]
  \centerline{\hfill
    \includegraphics[width=0.46\textwidth,clip]{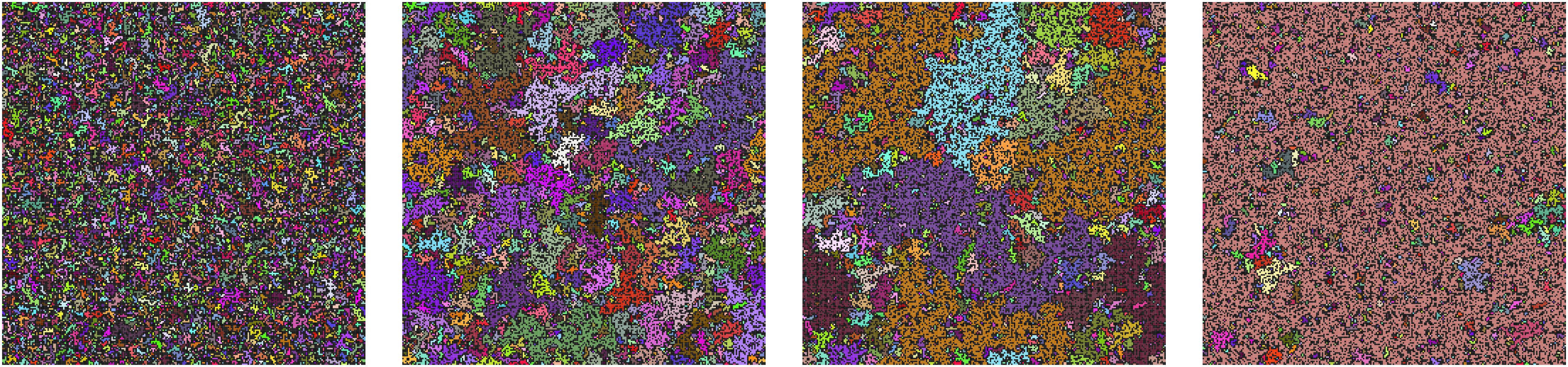}\hfill
    \vrule\hfill
    \includegraphics[width=0.46\textwidth,clip]{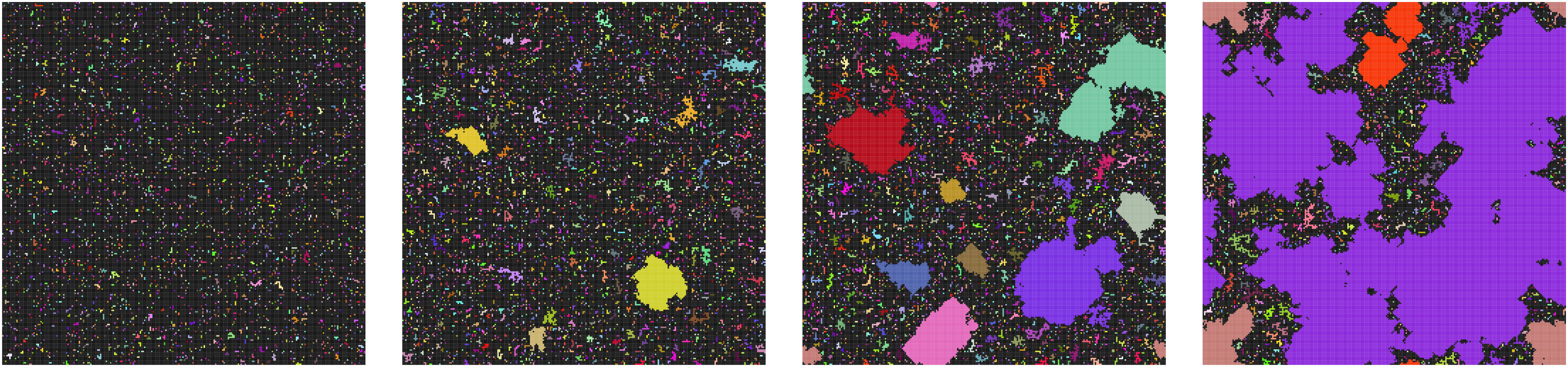}\hfill}
  \centerline{\hfill (a) \hfill \hfill (b) \hfill}
  \caption{\label{CrackGrowth} Growth of cracks for two different
    applied stresses (a) $f=0.1$ and (b) $f=0.4$ in a bundle of $256
    \times 256$ fibers at temperature $T=0.06$. The thresholds are
    generated according to Eq. \ref{eqPx} at low disorder
    $D=0.02$. The snapshots correspond to $t=0.40t_{\rm f}$,
    $0.60t_{\rm_f}$, $0.64t_{\rm f}$ and $0.68t_{\rm f}$ for $f=0.1$,
    and $t=0.48t_{\rm f}$, $0.70t_{\rm_f}$, $0.79t_{\rm f}$ and
    $0.87t_{\rm f}$ for $f=0.4$. The intact fibers are colored by
    black and the other colors represent different clusters of broken
    fibers. At low applied stress, cracks appear randomly in the
    system whereas at high applied stress localized cracks are
    observed to grow within the system.}
\end{figure}

The growth of cracks with time in the presence of temperature at two
different external stress $f=0.1$ and $f=0.4$ are shown in
Fig. \ref{CrackGrowth}. The intact clusters are colored by black and
the individual clusters of failed fibers are marked with different
colors. The pictures show two distinct regimes of crack growth. At low
applied stress the subsequent failure events appear randomly in space,
similar to the growth of percolation clusters. The clusters merge with
time and a spanning cluster appears in the system similar to a
percolating cluster that contains clusters of intact fibers of
different sizes inside it. On the other hand, at high stress,
localized clusters of failed fibers appear and that eventually
merge. The localized clusters in this case are compact.

\begin{figure}[t]
  \centerline{\hfill
    \includegraphics[width=0.32\textwidth,clip]{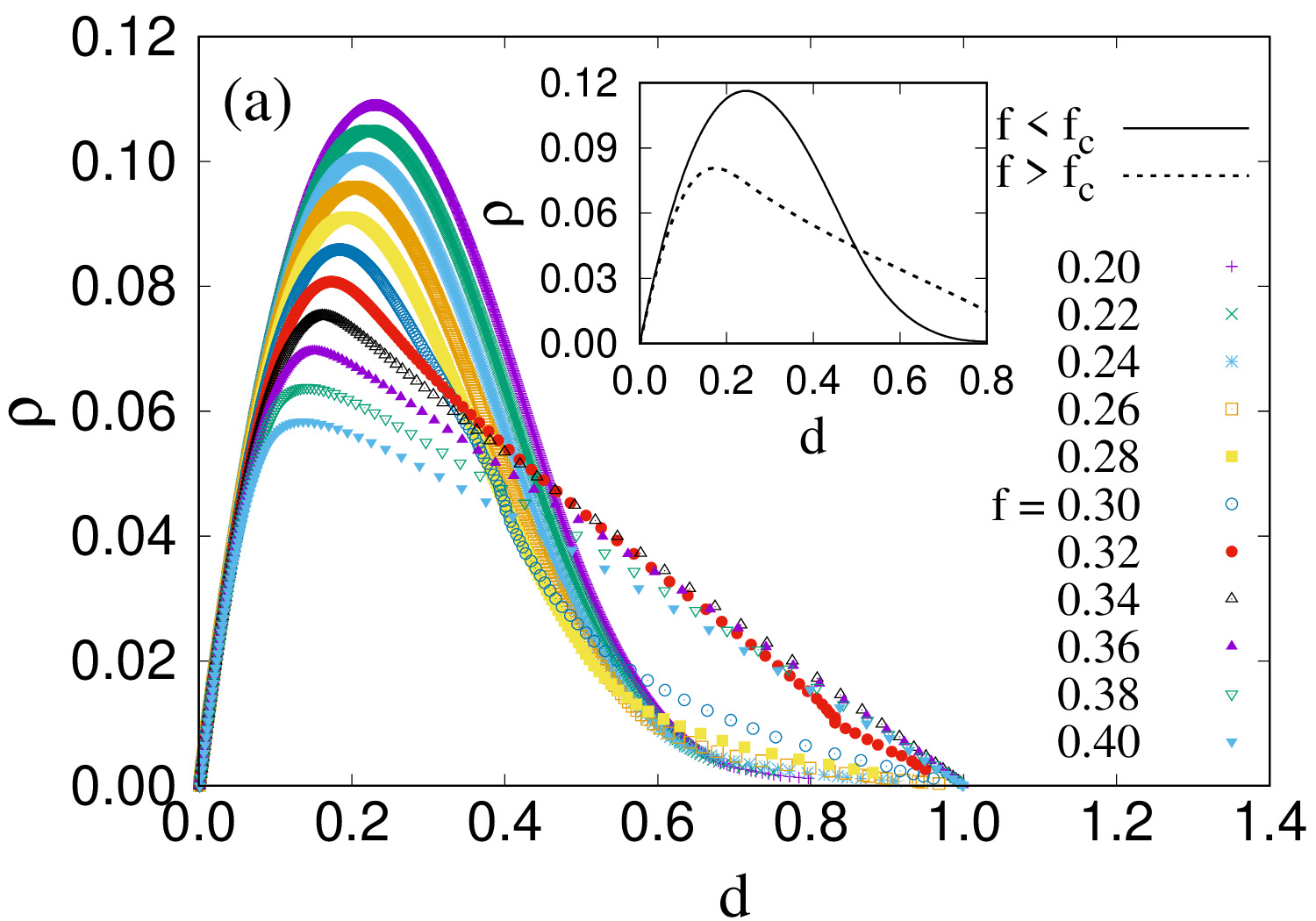}\hfill
    \includegraphics[width=0.32\textwidth,clip]{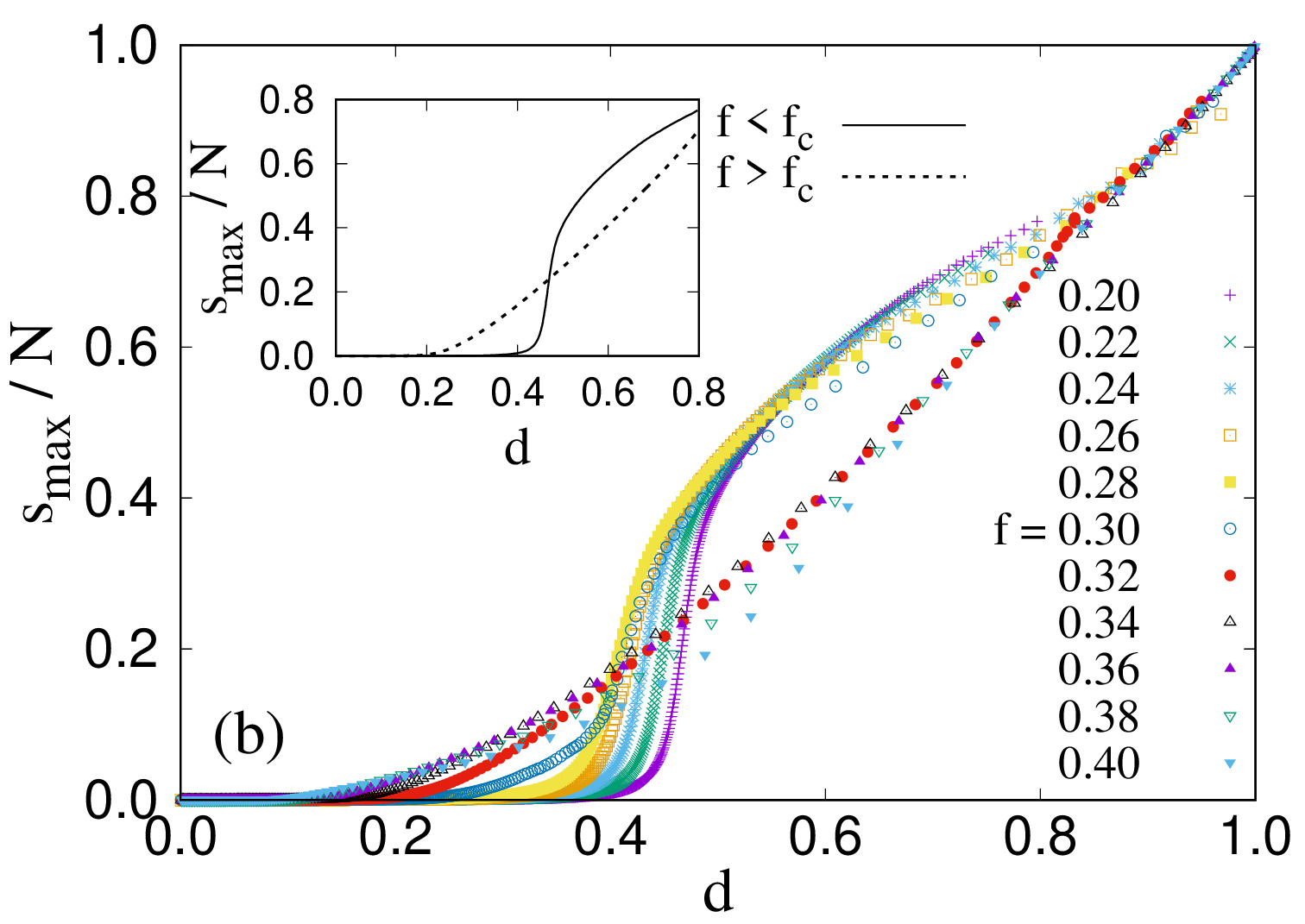}\hfill
    \includegraphics[width=0.32\textwidth,clip]{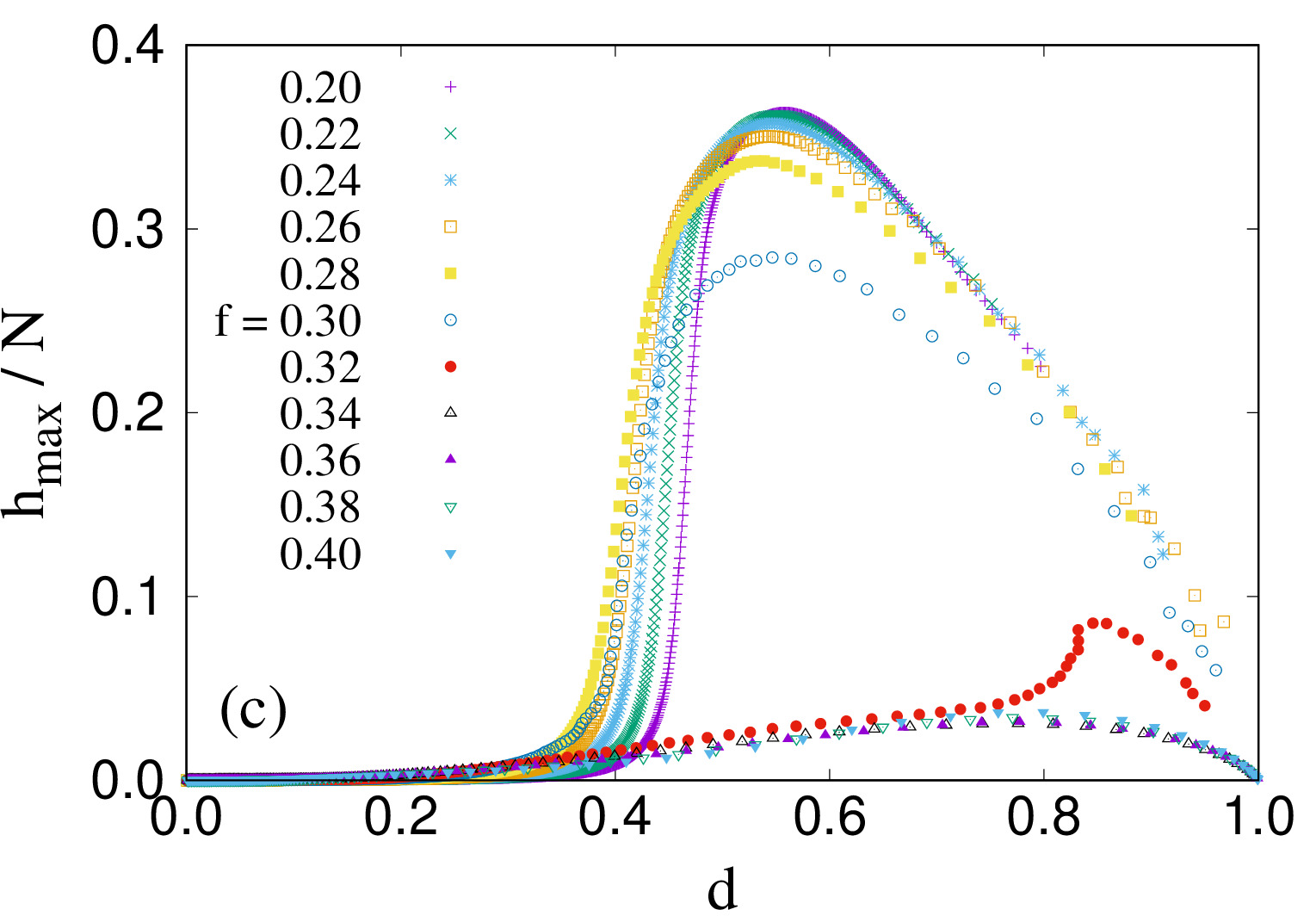}\hfill}
  \caption{\label{figGeo} (a) Variation of cluster density $\rho$ with
    damage $d$ for $T=0.06$ while varying the applied stress $f$. Two
    distinct behaviors are observed for the non-localized and
    localized regimes corresponding to $f<f_c$ and $f>f_c$
    respectively where $f_c=0.30$. These two behaviors are highlighted
    in the inset for $f=0.20$ and $0.32$. (b) Variation of the maximum
    crack size $s_{\rm max}$ normalized by the system size $N$, with
    damage $d$. The inset highlights the two different behaviors below
    and above $f_c$. (c) Variation of the normalized maximum hull size
    $h_{\rm max}/N$ with damage $d$. Notice that, the maxima of
    $h_{\rm max}$ for non-localized regime have much higher values
    compared to the localized regime indicating fractal structure of
    the clusters at $f<f_c$.}
\end{figure}

To study how the applied stress and temperature alter the failure
dynamics from random fractures to localized crack growth, we measure
three different geometrical quantities, the number density of
clusters, the largest cluster size and the largest perimeter size as a
function of damage $d$ for different values of $T$ and $f$. They are
plotted in Fig. \ref{figGeo} (a), (b) and (c) respectively. In
\ref{figGeo} (a), we plot the cluster number density $\rho$ as a
function of $d$ for different applied stress ($f$) at temperature
$T=0.06$. Here $\rho$ is defined as the number of clusters of failed
fibers divided by the total number of fibers $N$, which shows a
non-monotonic behavior with $d$. We observe $\rho=0$ at $d=0$ and
$\rho=1$ as $d\to 1$, because all the fibers are then broken creating
a single crack. In between, $\rho$ reaches a maximum value where the
system contains maximum number of clusters. This maximum decreases
with increasing $f$, and interestingly, the dependence of $\rho$ on
$d$ shows a characteristic difference beyond an applied stress $f>f_c$
indicating a difference in the failure dynamics. These two different
characteristics are highlighted in the inset for $f=0.20$ and $0.32$.

Next we measure the largest cluster size ($s_{\rm max}$) and the
largest perimeter size ($h_{\rm max}$) as a function of $d$. In
Fig. \ref{figGeo} (b) and (c), we plot $s_{\rm max}$ and $h_{\rm max}$
respectively, scaled by the number of fibers $N$. As we vary the
applied stress $f$, both the quantities show a characteristic
difference in the behavior below and above $f_c$. In the beginning of
the process, small fractures appear in the system leading to small
values of $s_{\rm max}$, which will eventually grow and merge with
each other. For $f>f_c$, the increase in $s_{\rm max}$ starts at a
much earlier point than that of $f<f_c$. This reflects the appearance
of compact localized cracks early in the system for $f>f_c$, which
grow more uniformly with time. Whereas for $f<f_c$, spatially
uncorrelated failures occur at random in space preventing the
appearance of a large cluster for a longer duration. At some point of
time, these small clusters start to coalesce causing a sharp increase
in $s_{\rm max}$ as seen in the figure. When the coalescence is over
after creating a large crack in the system, $s_{\rm max}$ grows in a
linear manner. Behavior of $s_{\rm max}$ shows similarity with that of
percolation model for $f<f_c$ whereas above $f_c$ it deviates from
such characteristic shape indicating the initiation of
localization. These two behaviors for $f<f_c$ and $f>f_c$ are
highlighted in the inset. The behavior of $h_{\rm max}$ shows a more
interesting picture. It shows two different behavior for $f<f_c$ and
$f>f_c$, however, the values of $h_{\rm max}$ are much smaller for
$f>f_c$, that is in the localized regime. This indicates fractal type
perimeter structure for $f<f_c$ compared to the perimeters of the
localized cracks for $f>f_c$. Later in section \ref{secBoun}, we will
use $h_{\rm max}$ to calculate the values of $f_c$ and will show how
the boundary between the non-localized and localized fracture regimes
vary with the temperature $T$. For $f<f_c$, both $s_{\rm max}$ and
$h_{\rm max}$ indicate percolation type dynamics during the failure
process which we will explore in detail in the following.

\begin{figure}[b]
  \centerline{\hfill
    \includegraphics[width=0.42\textwidth,clip]{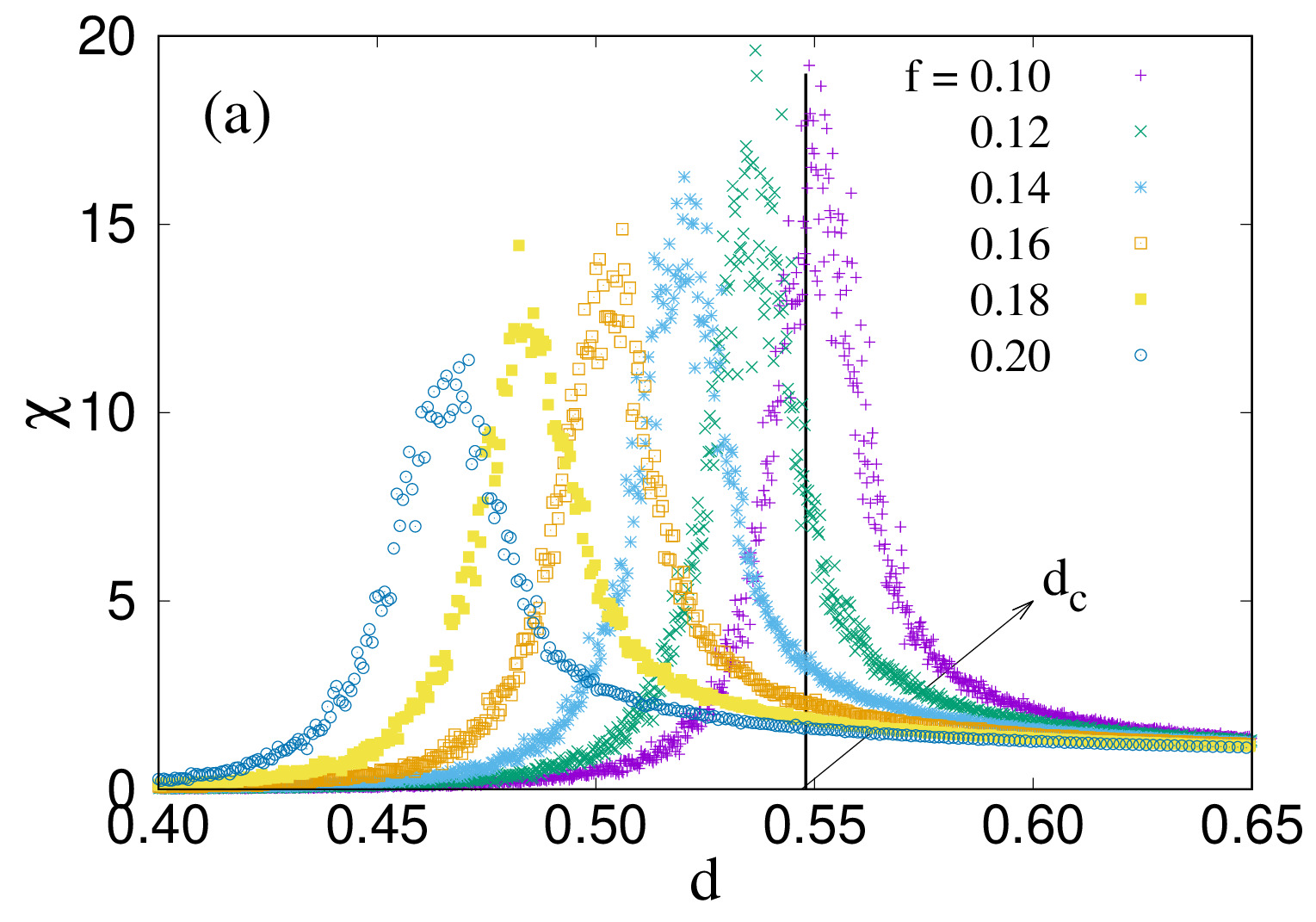}\hfill
    \includegraphics[width=0.42\textwidth,clip]{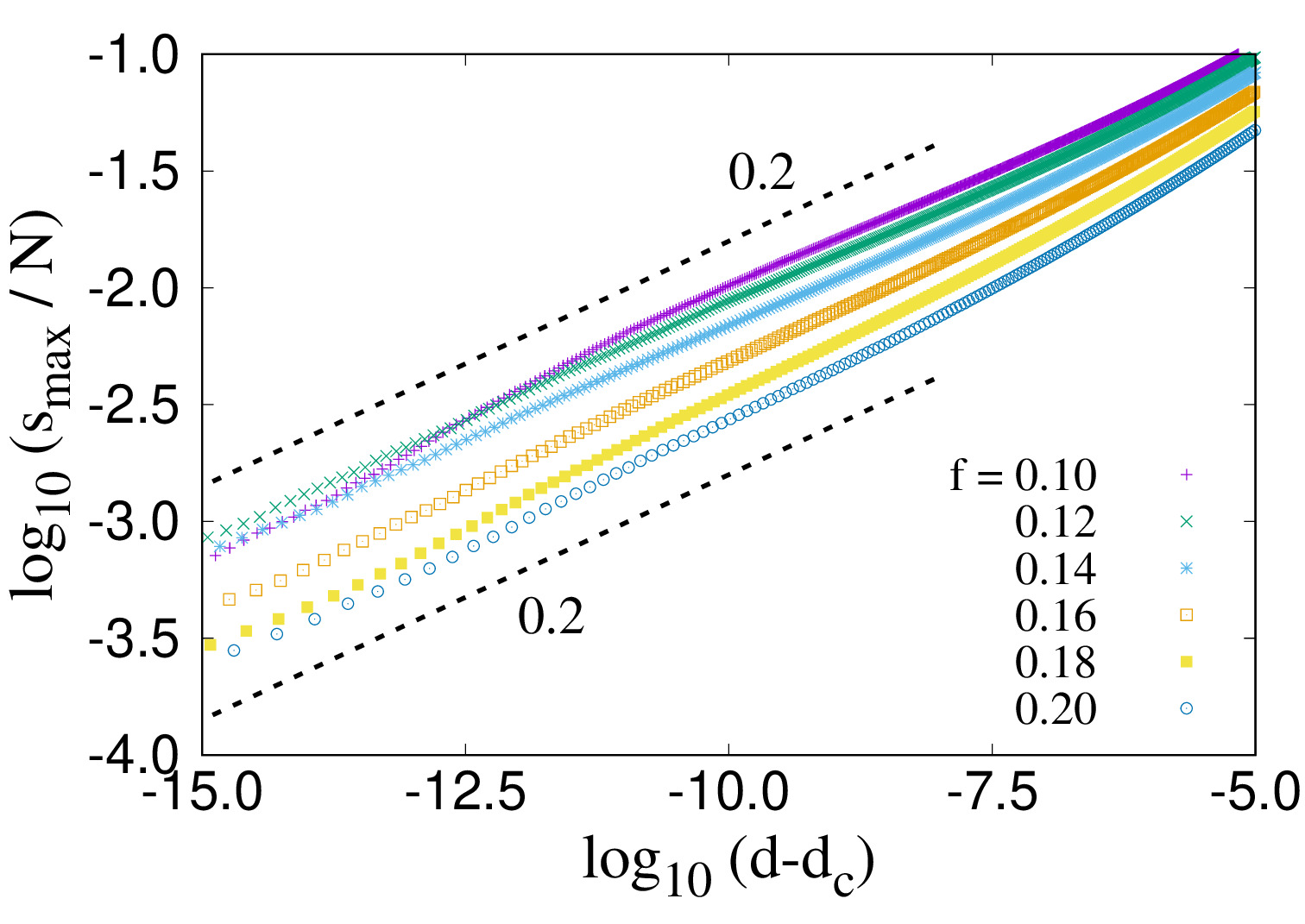}\hfill}
  \caption{\label{figCLMax} (a) The response function $\chi$, i.e.,
    the rate of change of largest cluster size $s_{\rm max}$ as a
    function of the damage $d$. This is obtained by taking the
    derivative of $s_{\max}$ with respect of $d$ using the central
    difference technique. From the peaks, we find the values of
    critical damage $d_c$. In (b), we plot $s_{\rm max}/N$ with
    $(d-d_c)$ where the slopes correspond to the exponent $\beta$
    defined in Eq. \ref{eqBeta}. We observe $\beta$ around $0.2$,
    which however fluctuates with the applied stress. The full list of
    the values of $\beta$ for different applied stress will be given
    in Tab. \ref{tabExp}.}
\end{figure}

\begin{figure}[t]
  \centerline{\hfill
    \includegraphics[width=0.42\textwidth,clip]{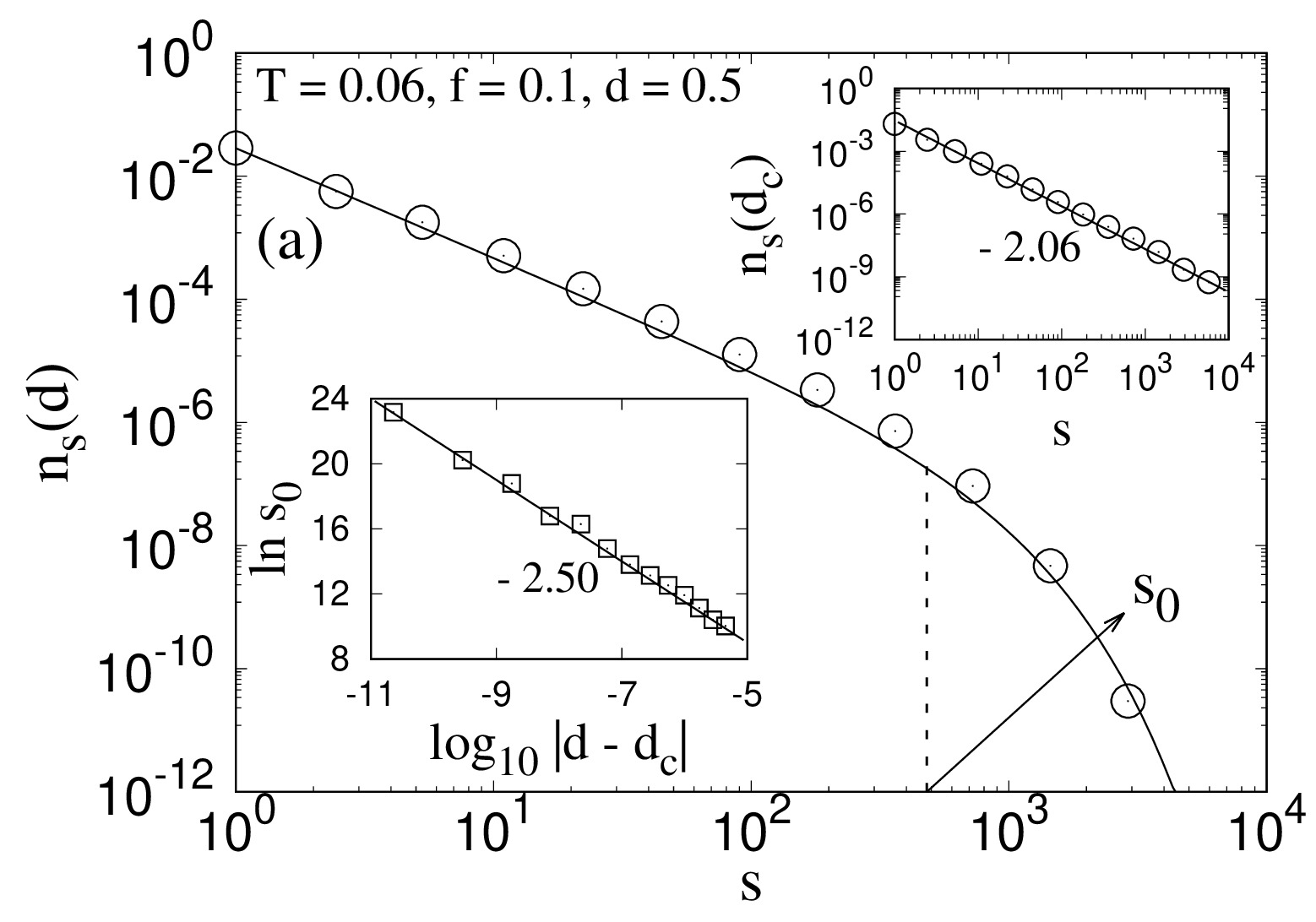}\hfill
    \includegraphics[width=0.42\textwidth,clip]{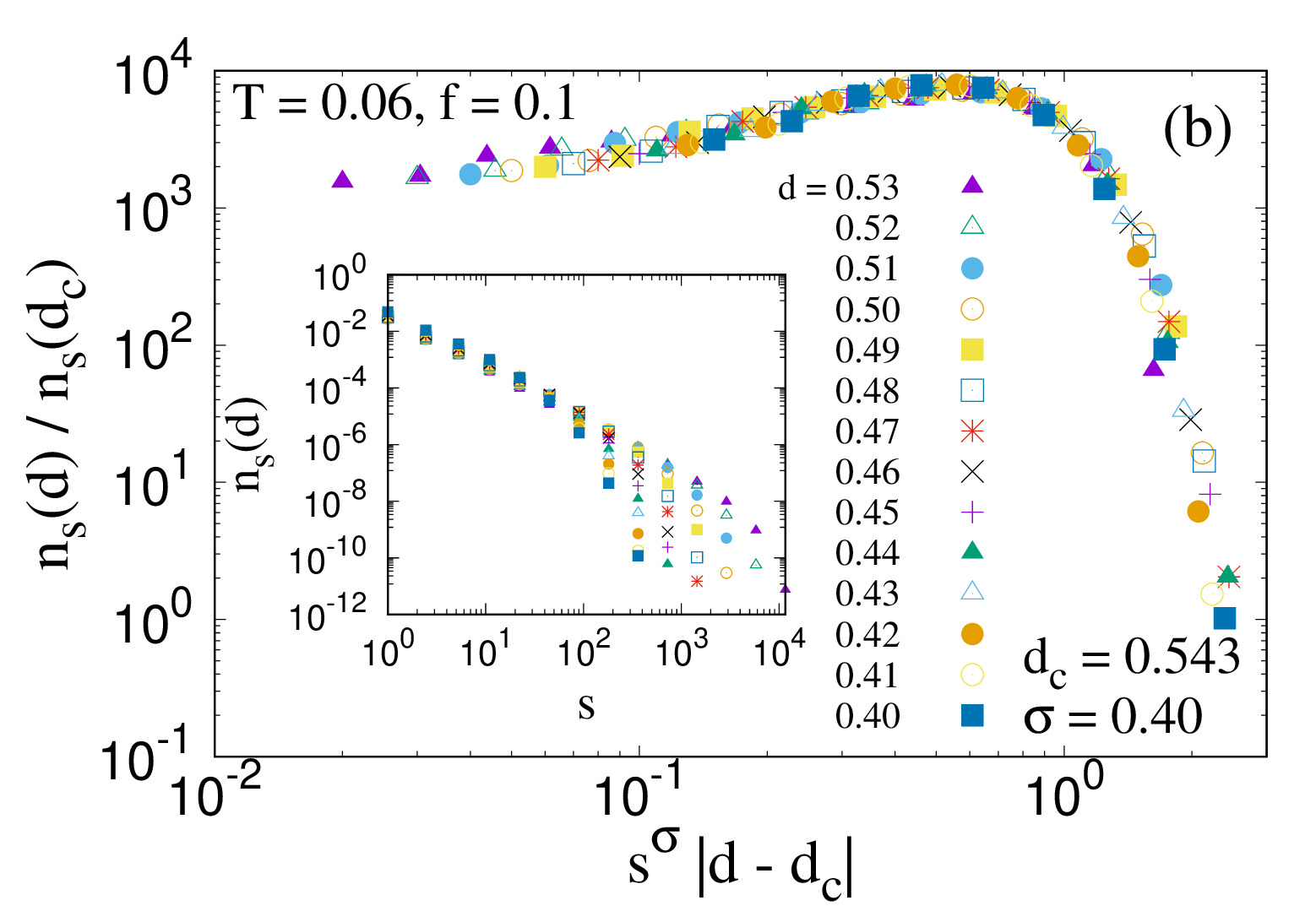}\hfill}
  \caption{\label{figDist} (a) Plot of cluster size distribution
    $n_s(d)$ at $d=0.5$ for $T=0.06$ and $f=0.10$. The circles
    represent simulation data and the solid line represents least
    square fitting of data with Eq. \ref{eqExp}, where $\tau$ is
    $2.06$. The value of $\tau$ is obtained from the distribution
    $n_s(d_c)$ using Eq. \ref{eqTau} as shown in the upper inset of
    (a). We obtain the values of $s_0$ from the least-square fitting
    for different values of $d$. The variation of $s_0$ with $|d-d_c|$
    is shown the inset, the slope of which corresponds to $1/\sigma$
    (Eq. \ref{eqS0}). In (b), we plot the scaled cluster size
    distribution $n_s(d)/n_s(d_c)$ as a function of the scaled
    variable $s^\sigma|d-d_c|$ for $T=0.06$ and $f=0.10$ where we have
    used $\sigma=1/2.50=0.40$ obtained from the previous plot. Here
    $d_c$ is $0.543$. In the inset, we show the unscaled behavior of
    $n_s(d)$ vs $s$ for different values of $d$.}
\end{figure}

\subsection{\bf Characterization of non-localized cluster growth for $f < f_c$}
To characterize the growth of fractures in the non-localized regime we
will now study the clusters geometries by using the percolation
framework \cite{s79,sa92}. We performed simulations at $T=0.06$ at
different values of $f<f_c$. We found $f_c\approx 0.3$ for $T=0.06$
and therefore considered $f=0.10$, $0.12$, $0.14$, $0.16$ and $0.18$
for these simulations. To find a percolation threshold or a critical
damage $d_c$, one can calculate the order parameter $P_\infty$ defined
as the probability that a site belongs to an infinite
cluster. $P_\infty$ can be measured from the number of sites ($s_{\rm
  max}$) that belong to the largest cluster and then by estimating
$P_\infty = \lim_{N\to\infty}P_N$ where $P_N=s_{\rm max}/N$ averaged
over all configurations. Ideally, $P_\infty$ approaches to $0$ from
$1$ continuously as $d\to d_c$ from above. In our case, we have a
finite system of $256^2$ fibers and therefore $P_N=s_{\max}/N$ does
not meet to $0$ sharply at a particular value of $d$
(Fig. \ref{figGeo} (b)). We therefore measure the percolation
threshold $d_c$ from the maximum of the response function $\chi$ of
$P_N$, that is when the rate of change of $s_{\rm max}$ with $d$ is
maximum. This is plotted in Fig. \ref{figCLMax} (a) where the peaks of
the plots determine the values of $d_c$ for different $f$. As $d\to
d_c$ from above, $P_\infty$ approaches to zero with a critical
exponent $\beta$ defined as,
\begin{align}
  \label{eqBeta}
  P_\infty \sim (d-d_c)^\beta\;.
\end{align}
In Fig. \ref{figCLMax}, we plot $s_{\rm max}/N$ as a function of
$(d-d_c)$ for different values of $f$ and from the slopes we find
$\beta$ in the range of $0.18$ to $0.23$ with error bars ranging
between $0.04$ to $0.07$ (see Tab. \ref{tabExp}). For site
percolation, $\beta=5/36$.

Next, we study the distribution of clusters of failed fibers near the
critical damage $d_c$ and investigate the scaling behavior of its
moments. The cluster size distribution function $n_s(d)$ is defined as
the number of $s$-sized finite clusters per lattice site at a damage
$d\le d_c$. According to scaling hypothesis \cite{s79,sa92}, the
functional form of the distribution can be assumed as,
\begin{align}
  \label{eqNs1}
  \displaystyle
  \frac{n_s(d)}{n_s(d_c)} \sim \Phi\left[s^{\sigma}|d-d_c|\right]\;,
\end{align}
where $n_s(d_c)$ is the cluster size distribution at the critical
damage $d_c$ and $\sigma$ is a critical exponent. This $n_s(d_c)$ at
the critical damage shows a power-law decay,
\begin{align}
  \label{eqTau}
  n_s(d_c) \sim s^{-\tau}\;,
\end{align} 
where $\tau$ is the cluster size distribution exponent. Using this in
Eq. \ref{eqNs1}, we have the functional form for $n_s(d_c)$,
\begin{align}
  \label{eqNs2}
  n_s(d) \sim s^{-\tau} \Phi\left[s^{\sigma}|d-d_c|\right]
\end{align}
as $d\to d_c$. Away from the critical damage, $n_s(d_c)$ falls with an
exponential cut off with the following form,
\begin{align}
  \label{eqExp}
  n_s(d) \sim s^{-\tau} \exp\left(-\displaystyle\frac{s}{s_0}\right)\;,
\end{align} 
where $s_0$ is the exponential cut off. As $d\to d_c$, $s_0$ generally
shows the power law dependency,
\begin{equation}
  \label{eqS0}
  s_0 \sim |d-d_c|^{-1/\sigma}\;.
\end{equation}
To verify the scaling functional form for the cluster size
distribution, we first measure the distribution of clusters at the
critical point $d_c$. We then estimate the exponent $\tau$ by using
Eq. \ref{eqTau} as shown in the upper inset of Fig. \ref{figDist} (a).
Using the value of $\tau$, we plot in Fig. \ref{figDist} (a) $n_s(d)$
as a function of $s$ for $d=0.5$. There we used least square fitting
of the data points with the functional form given in Eq. \ref{eqExp}
and estimated $s_0$ for different values of $d$. We then plot $s_0$ as
a function of $|d-d_c|$ as shown in the lower inset of
Fig. \ref{figDist} (a) and determine the exponent $\sigma=0.40$ from
the slope (Eq. \ref{eqS0}). With this value of $\sigma$, we verify the
scaling function form given in Eqs. \ref{eqNs1}. In Fig. \ref{figDist}
(b), we plot the scaled cluster size distribution $n_s(d)/n_s(d_c)$ as
a function of the scaled variable $s^\sigma|d-d_c|$ for $T=0.06$ and
$f=0.10$ where $\sigma$ is taken as $0.40$. We observe a data collapse
for different values of $|d-d_c|$ and $s$ showing the validity of
Eq. \ref{eqNs1}. In the inset, we plot the unscaled values of $n_s(d)$
which show the approach of the power law behavior given in
Eq. \ref{eqNs2} as $d\to d_c$.

\begin{figure}[t]
  \centerline{\hfill\includegraphics[width=0.42\textwidth,clip]{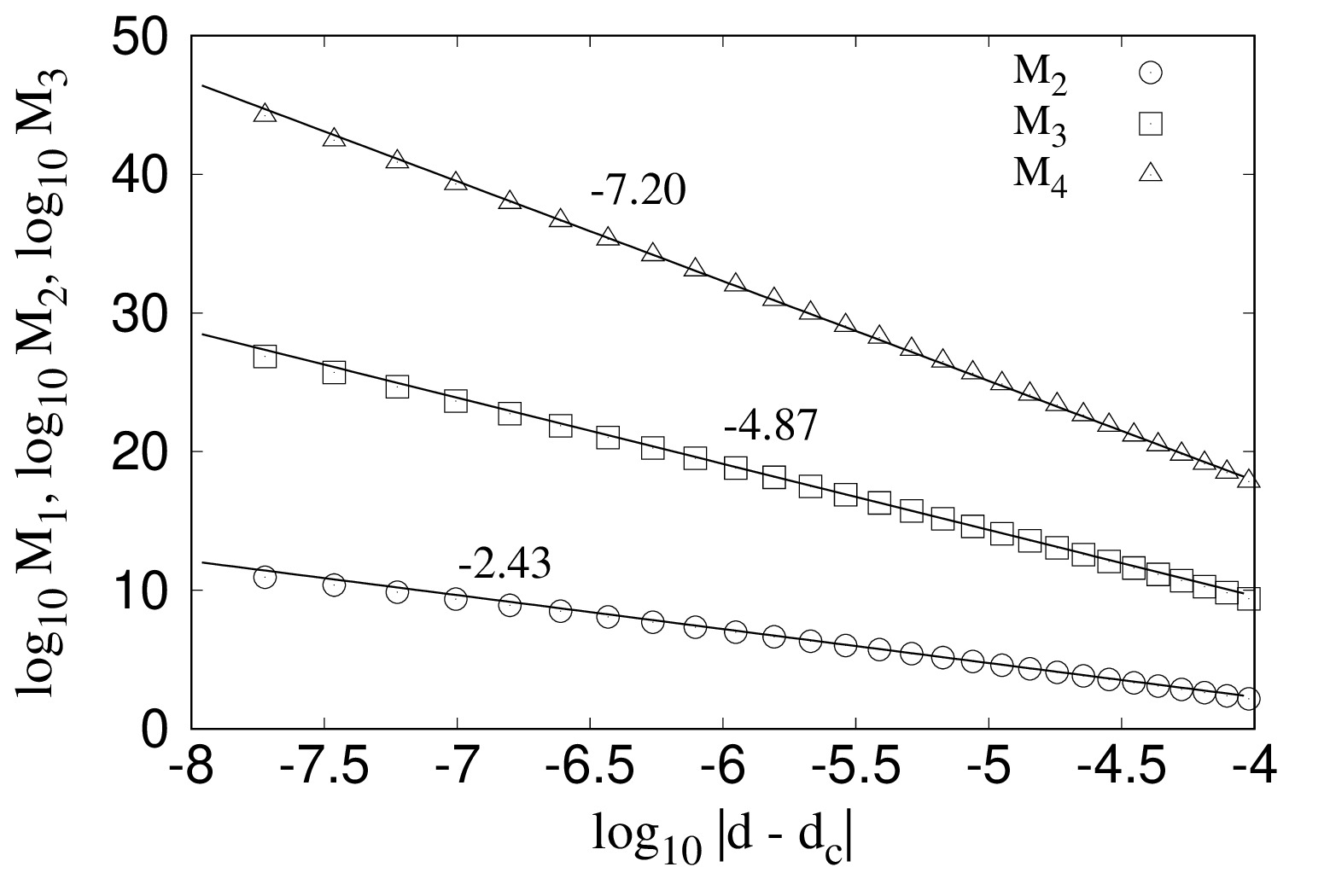}\hfill}
  \caption{\label{figMom} Plot of the second, third and fourth moments
    $M_2$, $M_3$ and $M_4$ of cluster size distribution function as a
    function of $|d-d_c|$ for $T=0.06$ and $f=0.10$. From the slopes,
    we obtain the exponents, $\gamma=2.43 \pm 0.03$, $\delta=4.87 \pm
    0.05$ and $\eta=7.20 \pm 0.07$.}
\end{figure}

The $k$th {\it moment} of the cluster-size distribution, $M_k =
\sum_s^\prime s^kn_s(d)$ shows the singularity,
\begin{align}
  \label{eqMom1}
  M_k \sim |d-d_c|^{(\tau-1-k)/\sigma}
\end{align} 
as $d\to d_c$. The primed summation for $M_k$ indicates the summation
over finite clusters. The first moment, $k=1$ represents the
percolation probability $P_\infty$. We measure the next three moments
for $k=2$, $3$ and $4$, where $k=2$ corresponds to the average cluster
size. As $d \rightarrow d_c$, the moments $M_2$, $M_3$ and $M_4$
diverge with their respective critical exponents $\gamma$, $\delta$
and $\eta$ defined as,
\begin{equation}
  \label{eqMom2}
  \displaystyle
  M_2\sim|d-d_c|^{-\gamma} \quad,\qquad M_3\sim|d-d_c|^{-\delta} \qquad{\rm and}\qquad M_4\sim|d-d_c|^{-\eta}\;.
\end{equation}
Using Eq. \ref{eqMom1}, we have the scaling relations connecting the
moment exponents with $\tau$ and $\sigma$,
\begin{equation}
  \label{eqScl1}
  \beta=(\tau-2)/\sigma \quad,\qquad \gamma=(3-\tau)/\sigma \quad,\qquad \delta=(4-\tau)/\sigma \qquad{\rm and}\qquad \eta=(5-\tau)/\sigma\;.
\end{equation}
Furthermore, we can eliminate $\tau$ and $\sigma$ to find the scaling
relations between the moment exponents,
\begin{equation}
  \label{eqScl2}
  \delta=\beta+2\gamma \qquad{\rm and}\qquad \eta=2\delta-\gamma\;.
\end{equation}
In Fig. \ref{figMom}, we plot the moments for $T=0.06$ and
$f=0.10$. From the slopes, we find the values of the exponents,
$\gamma=2.43 \pm 0.03$, $\delta=4.87 \pm 0.05$ and $\eta=7.20 \pm
0.07$. The values are close to those of site percolation model in
2D. The exponents satisfy the relationship given in Eqs. \ref{eqScl1}
and \ref{eqScl2} within error bar. The full list of exponents for
different values of the applied stress $f$ is listed in Table
\ref{tabExp} where the verification of the scaling relations in
Eqs. \ref{eqScl1} and \ref{eqScl2} are also indicated.

\begin{table}[b]
  \centering
  \begin{tabular}{|c|c|c|c|c|c|} \hline
    Applied stress ($f$)        & $0.10$           & $0.12$           & $0.14$           & $0.16$           & $0.18$           \\ \hline
    $d_c$                       & $0.543\pm 0.001$ & $0.537\pm 0.001$ & $0.522\pm 0.001$ & $0.503\pm 0.001$ & $0.485\pm 0.001$ \\ \hline
    $\beta$                     & $0.20 \pm 0.05$  & $0.19 \pm 0.06$  & $0.21 \pm 0.04$  & $0.23 \pm 0.07$  & $0.18 \pm 0.05$  \\ \hline
    $\gamma$                    & $2.43 \pm 0.03$  & $2.40 \pm 0.04$  & $2.39 \pm 0.03$  & $2.45 \pm 0.02$  & $2.44 \pm 0.02$  \\ \hline
    $\delta$                    & $4.87 \pm 0.05$  & $4.83 \pm 0.04$  & $4.89 \pm 0.03$  & $4.88 \pm 0.02$  & $4.92 \pm 0.02$  \\ \hline
    $\eta$                      & $7.20 \pm 0.07$  & $7.23 \pm 0.09$  & $7.15 \pm 0.08$  & $7.20 \pm 0.09$  & $7.19 \pm 0.07$  \\ \hline
    $\tau$                      & $2.06 \pm 0.05$  & $2.03 \pm 0.04$  & $2.07 \pm 0.04$  & $2.06 \pm 0.02$  & $2.10 \pm 0.03$  \\ \hline
    $\sigma$                    & $0.40 \pm 0.02$  & $0.39 \pm 0.04$  & $0.43 \pm 0.03$  & $0.41 \pm 0.02$  & $0.38 \pm 0.03$  \\ \hline
    $\delta$ (Eq. \ref{eqScl2}) & $5.06 \pm 0.11$  & $4.99 \pm 0.14$  & $4.99 \pm 0.10$  & $5.13 \pm 0.11$  & $5.06 \pm 0.09$  \\ \hline
    $\eta$   (Eq. \ref{eqScl2}) & $7.31 \pm 0.13$  & $7.26 \pm 0.12$  & $7.39 \pm 0.09$  & $7.31 \pm 0.06$  & $7.40 \pm 0.06$  \\ \hline
    $\sigma$ (Eq. \ref{eqScl1}) & $0.37 \pm 0.12$  & $0.34 \pm 0.16$  & $0.38 \pm 0.12$  & $0.36 \pm 0.14$  & $0.43 \pm 0.15$  \\ \hline
  \end{tabular}
  \caption{\label{tabExp} List of the threshold damage and critical
    exponents obtained for different applied stress in the
    non-localized regime. Exponents are close to those observed in 2D
    percolation model where $\beta=5/36$, $\gamma=43/18$,
    $\tau=187/91$ and $\sigma=36/91$. We also show in the bottom three
    rows the calculation of $\eta$, $\delta$ and $\sigma$ by using the
    scaling relations in Eqs. \ref{eqScl1} and \ref{eqScl2} which can
    be compared with the measured values of the exponents.}
\end{table}

\begin{figure}[t]
  \centerline{\hfill\includegraphics[width=0.42\textwidth,clip]{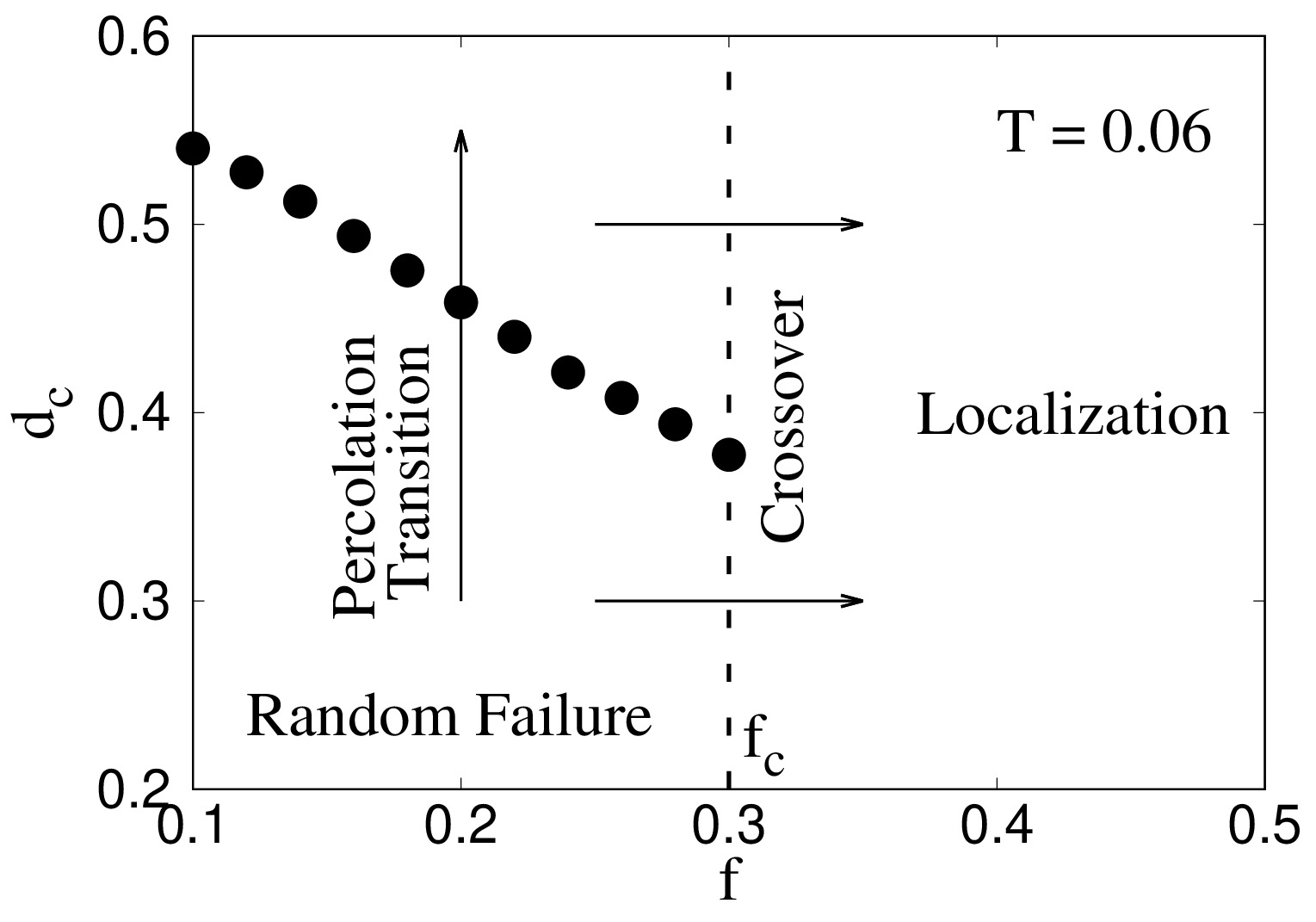}\hfill}
  \caption{\label{figDcF} The figure shows two different regions: (I)
    random failure and (II) localization, depending on whether the
    stress applied on the model is below or above $f_t$ (the crossover
    point). For $f>f_c$, individual patches grows within the
    bundle. For $f<f_c$, the failure process is random and the model
    undergoes a percolation transition around damage $d_c$. We keep
    $T=0.06$ and $D=0.02$.}
\end{figure}

\begin{figure}[h]
  \centerline{\hfill
      \includegraphics[width=0.32\textwidth,clip]{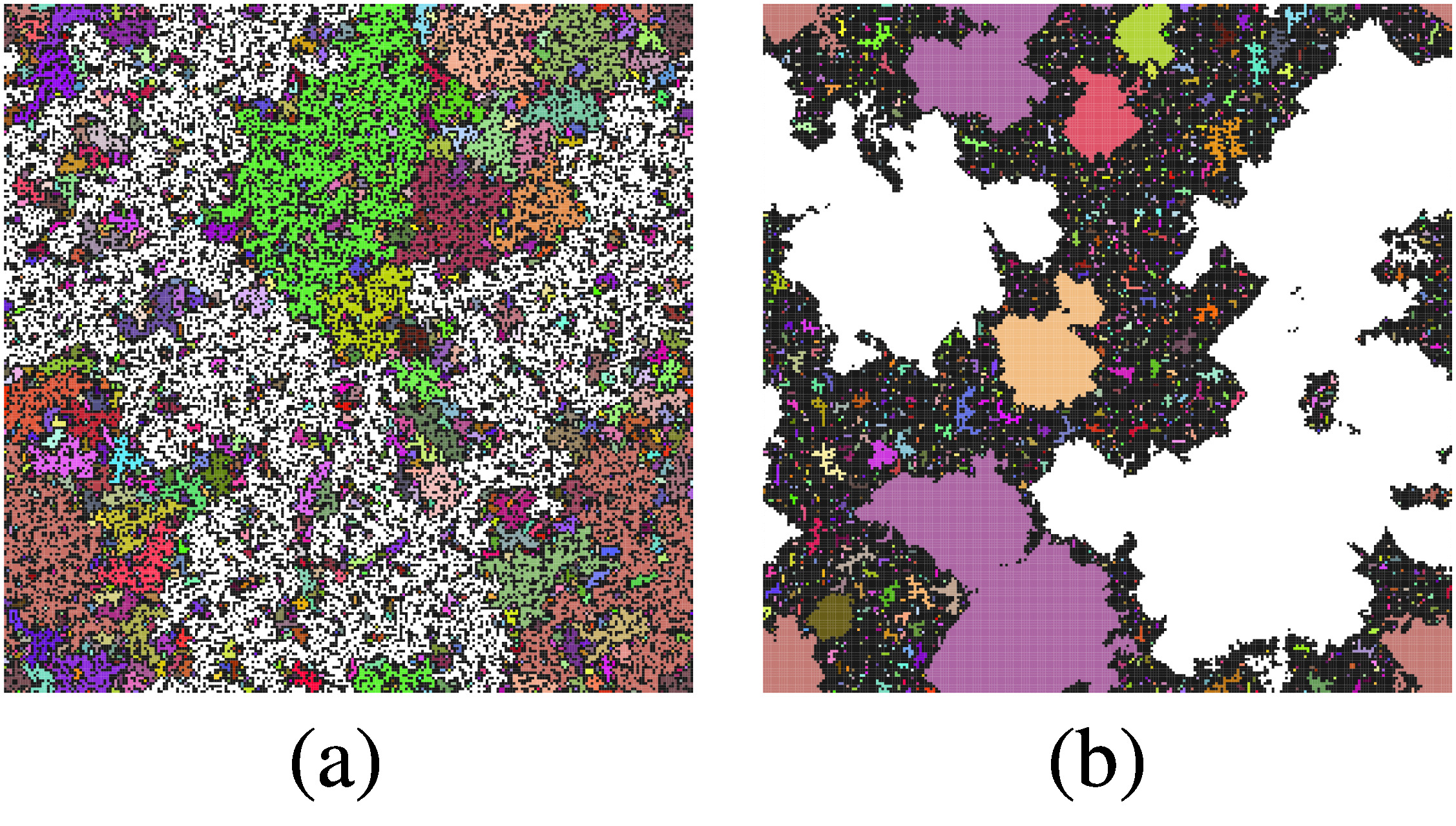}\hfill \hfill
      \includegraphics[width=0.32\textwidth,clip]{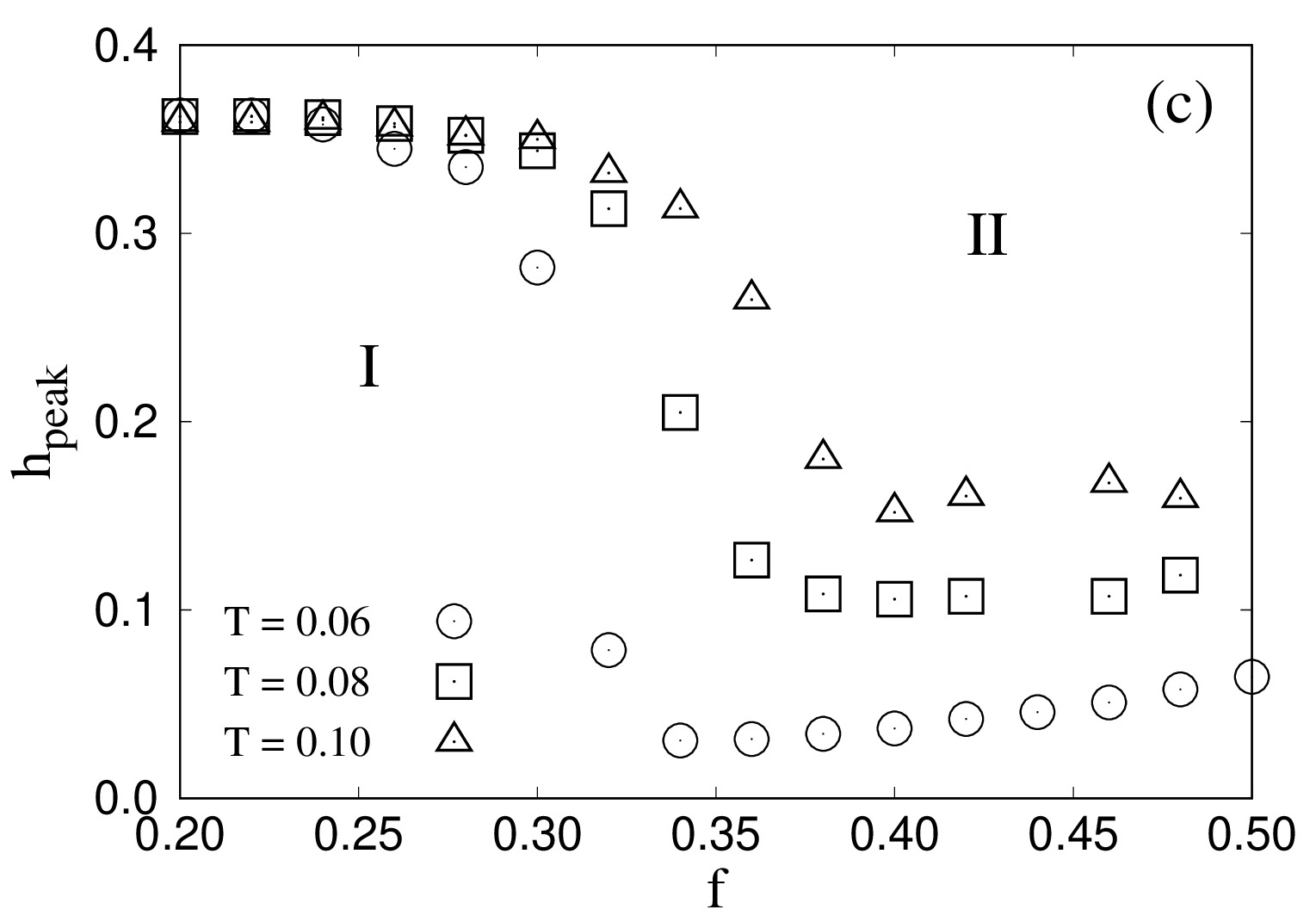}\hfill
      \includegraphics[width=0.32\textwidth,clip]{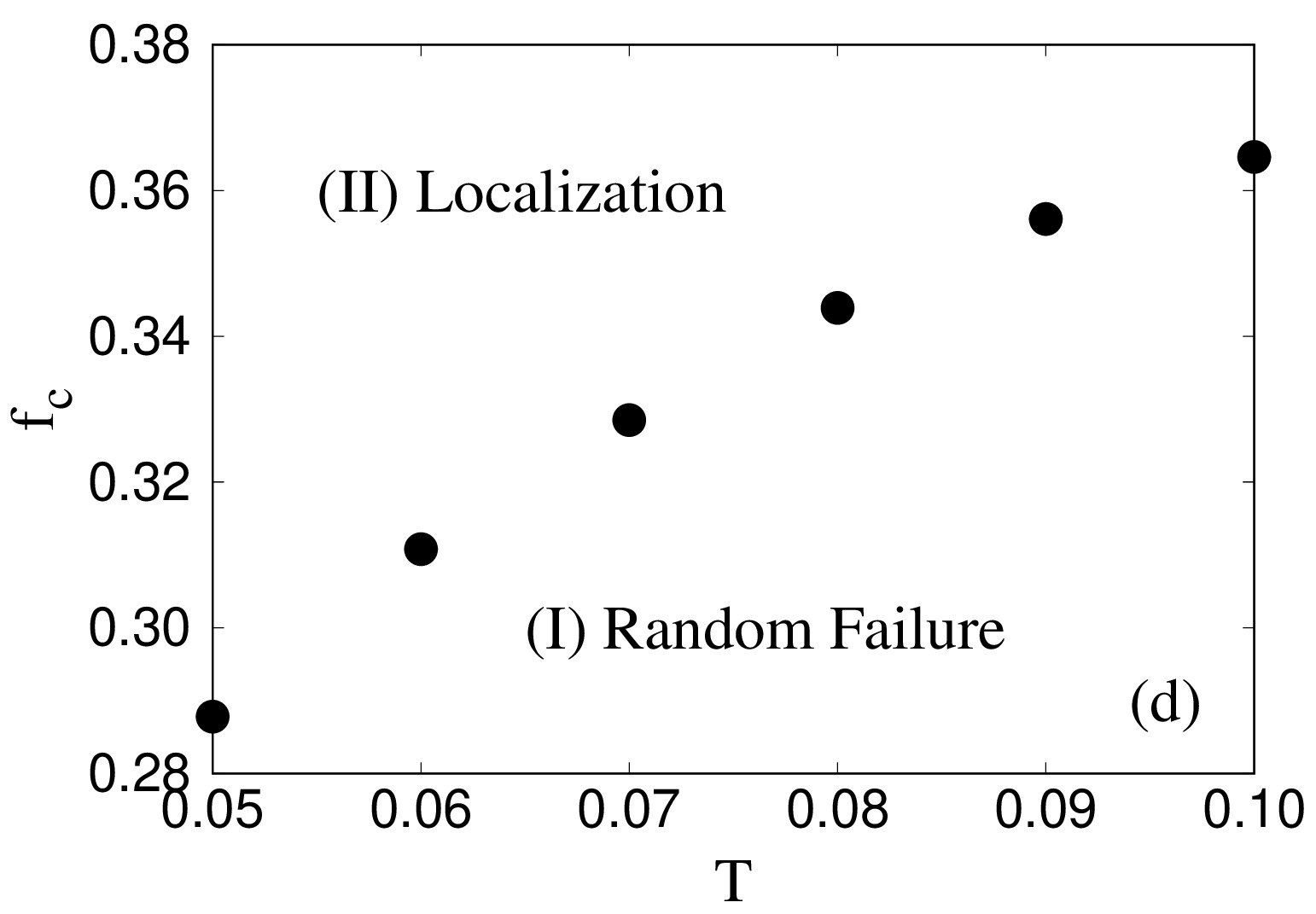}\hfill}
  \caption{\label{figHMax} Snapshots of the system for (a)
    non-localized ($f=0.1$) and (b) localized ($f=0.4$) growths when a
    spanning cluster appears in the system. The largest cluster is
    marked by white. In (c), we show the variation of $h_{\rm peak}$
    with applied stress $f$ where $h_{\rm peak}$ has a high value at
    low applied stress. As $f$ is increased beyond a threshold value
    $f_c$, $h_{\rm peak}$ decreases rapidly and saturates at
    sufficiently low value. In (d) we show the boundary separating two
    different regimes, (I) the percolation type random failure and
    (II) the localized crack growth. As temperature increases, the
    localization takes place at a relatively higher applied stress as
    the stress concentration has to overcome a larger thermal
    randomness.}
\end{figure}

\subsection{\bf Boundary between the random and localized failure}
\label{secBoun}
We show in Fig. \ref{figDcF}, the variation of the critical damage
$d_c$ with the applied stress $f$ for $T=0.06$. The system undergoes a
percolation transition at $d_c$. The value of $d_c$ decreases with the
increase in $f$ for $f<f_c$. Above $f_c$, there is a crossover in the
failure process from a percolation type growth to a compact localized
growth where the geometrical quantities show different behavior as
observed in Fig. \ref{figGeo}. To find the boundary that separates the
random failure from the localized fracture growth as a function of
temperature, we focus on the largest perimeter size $h_{\rm max}$
which shows abrupt decrease in the peak value when the applied stress
crosses $f_c$. As mentioned before, this is due to the fractal
structure of the percolation clusters at $d_c$ which makes the hull
highly rarefied and long compared to the perimeters of the compact
clusters in the localized growth. Two such clusters are shown in
Fig. \ref{figHMax} (a) and (b) for the percolation type growth and the
localized growth respectively where the largest clusters are colored
by white. In Fig. \ref{figHMax} (c), we plot $h_{\rm peak}$, the
maximum value of the largest perimeter size $h_{\rm max}$ during a
failure process, as a function of the applied stress $f$ for three
different temperatures $T=0.06$, $0.08$ and $0.10$. The plots show two
distinct regimes, a regime (I) with fairly constant higher values,
then another regime (II) where $h_{\rm peak}$ falls rapidly to a
relatively lower value as $f$ crosses $f_c$. By measuring the values
of $f_c$ for different temperatures, we find the boundary on the $f_c$
vs $T$ plane that separates the random failure (I) from the localized
failure process (II). As temperature is increased, a higher applied
stress is required to establish localization in the model. This is
because the increase in temperature increases thermal fluctuations
that lead to higher spatial randomness in the failure events. In this
case, a large amount of stress localization is required to counteract
the thermal fluctuation.

\section{Conclusions}
\label{secCon}
We have included thermal noise in an LLS fiber bundle model to study
creep and subsequent failure through the interplay between thermal
fluctuations and stress concentration due to externally applied
stress. We observed that the presence of thermal fluctuations makes
the failure events spatially uncorrelated and non-localized even if
the strength of system disorder is low. This non-localized fracture
growth shows a percolation transition governed by critical exponents
and scaling relations. In the absence of temperature, the same amount
of system disorder shows a highly correlated failure process in LLS
fiber bundle model \cite{srh20}. Such spatial correlation can be
obtained in presence of temperature as well when the applied stress is
sufficiently high so that the effect of stress localization can outrun
the effect of thermal fluctuations. This is a new mechanism for
establishing localization. Stormo et. al \cite{sgh12} observed
localized failure in a soft clamp model \cite{bhs02} as the elasticity
of the clamps are decreased. In fiber bundle model the disorder
strength and the stress release range plays a crucial role in
determining the correlations in space. It was recently shown
\cite{srh20} that as the strength of disorder is tuned, a percolation
transition is observed from a localized phase to a non-localized
phase. One can also approach a random failure process if the stress
release range is increased \cite{brr15,rbr17} instead of the disorder
strength. In the present study, we show that as the temperature is
increased, a larger stress has to be applied to overcome the thermal
fluctuations and to initiate the localization. In this work, however,
we have not addressed the combined effect of varying both the thermal
noise and system disorder during the failure process, which will be
interesting to study in the future.

\section*{Acknowledgment}
\label{Acknow}
This work was partly supported by the Research Council of Norway
through its Centers of Excellence funding scheme, project number
262644 and by the National Natural Science Foundation of China under
grant number 11750110430.

\end{document}